\documentclass[aps, twocolumn, reprint, amsmath, floatfix, superscriptaddress]{revtex4-1}

\usepackage{graphicx} 
\graphicspath{{./figures/}} 

\usepackage{comment} 
\usepackage{dcolumn} 
\usepackage{bm} 
\usepackage{parskip} 

\usepackage{natbib} 
\usepackage{hyperref} 
\usepackage{footnote}
\usepackage{xcolor}
\usepackage{amsfonts}

\begin{document}
\preprint{APS/123-QED}
\title{Signatures of directed and spontaneous flocking}

\author{Martino Brambati}
\affiliation{Dipartimento di Scienza e Alta Tecnologia, Universit{\'a} degli Studi dell'Insubria, Como, Italy}

\author{Giuseppe Fava}
\author{Francesco Ginelli}
\affiliation{Dipartimento di Scienza e Alta Tecnologia and Center for Nonlinear and Complex Systems, Universit{\'a} degli Studi dell'Insubria, Como, Italy}
\affiliation{INFN sezione di Milano, Milano, Italy}

\date{\today}
\begin{abstract}
Collective motion — or {\it flocking} -- is an emergent phenomena that underlies many biological processes of relevance, from cellular migrations to animal groups movement.  In this work, we derive scaling relations for the fluctuations of the mean direction of motion and for the static density structure factor (which encodes static density fluctuations) in the presence of a homogeneous, small external field. This allows us to formulate two different and complementary criteria capable of detecting instances of directed motion exclusively from easily measurable dynamical and static signatures of the collective dynamics,  without the need to detect correlations with environmental cues. The {\it static} one is informative in large enough systems, while the {\it dynamical} one requires large observation times to be effective. We believe these criteria may prove useful to detect or confirm the directed nature of collective motion in {\it in vivo} experimental observations, which are typically conducted in complex and not fully controlled environments. 
\end{abstract}

\maketitle
\section{\label{introduction}Introduction}
Collective motion is an ubiquitous emergent phenomena observed in a wide array of different living systems and on an even wider range of scales. Examples range from fish schools and flocks of birds to bacteria colonies and cellular migrations \cite{ramaswamy2010mechanics}. Its importance cannot be underestimated: cellular migration, for instance,  is a key driver of embryonic development, wound healing, and some types of cancer invasion \cite{Trepat2020}. Animal group movement may reduce risk of predation and underlies many foraging and migratory processes \cite{Krause2002}.

Collective motion -- or {\it flocking} -- involves a large number of active units, capable of self-propelled motion, that mutually synchronize their direction of motion. By doing so, they break their (continuous) rotational symmetry selecting a well defined mean direction for the entire group. This symmetry breaking process may be {\it spontaneous}, so that the emerging direction of motion is chosen by chance among the infinitely many available ones, as it has been observed in starling flocks \cite{cavagna2010scale}. But symmetry breaking may also be directed by  external cues or by some {\it a priori} knowledge of the group target (e.g.: a known foraging site). Cell motility, for instance, is known to be sensitive to a wide range of external gradients of chemical (chemotaxis), mechanical (durotaxis) and electrical (electrotaxis) origin \cite{directedcellmigration}.

A natural question when observing flocking phenomena, such as cellular migration or the coordinated movement of animal groups, thus regards the nature of collective motion. Is it a spontaneous phenomena, exclusively driven by the interactions between individual active units, or the observed group movement is being directed by some external factor?\\
While this distinction can be simply made {\it in vitro}, where experimental factors are easily controlled, {\it in vivo} observations are typically conducted in rather complex environments, making the task much more arduous \cite{VitrotoVivo}. Chemotactic guidance, for instance, is known to be involved in many instances of embryonic development \cite{embryogenesis}, but {\it in vivo} chemical or mechanical gradients have not systematically been observed for all cellular migration phenomena, leading authors to speculate about other types of spatial guidance clues which may be at play in certain instances \cite{voituriez2015}. 

It may thus be desirable to formulate simple criteria capable of discriminating in the first place between spontaneous collective motion, taking place in an isotropic environment and directed one, simply by observing the static and dynamical features of the active units involved, without the need to detect and establish correlations (or the lack of) with gradients or other environmental cues. Intuitively, the simplest signature of such a difference should lie in the persistence of the mean direction of motion, which is expected to be lower for spontaneous flocking. However -- as we will show in the following -- this simple criteria may fail for short observation times and/or small environmental anisotropies. An alternative approach we propose involves the observation of large wavelength {\it static} fluctuations in the active particles density, which are encoded by the density structure factor $S(q)$. Indeed, spontaneous collective motion implies a diverging structure factor at small wavenumbers $q$ \cite{toner1995long, toner1998flocks}, as experimentally confirmed in {\it in vitro} experiments of cellular migration \cite{giavazzi2017giant}. Here, we argue that such a divergence is suppressed in directed collective motion and that this fact can be used to successfully detect directed motion even on short or instantaneous observation timescales.\\

In this paper we discuss collective motion in the presence of  static and homogeneous small external fields of amplitude $h$ (i.e. in a {\it linear response} regime), showing that directed motion can be inferred from the fluctuations of the mean group direction (our so-called {\it dynamical approach}) only for observation times $t > \tau_c \sim h^{-1}$. On the contrary, the study of the system density fluctuations in large enough systems, may reveal directed motion also for much shorter observation times. In particular, we show that the free system structure factor's divergence is suppressed for wavenumbers $q<q_c \sim h^{1/z}$, where $z$ is the dynamical scaling exponent of the celebrated Toner \& Tu theory of Flocking \cite{toner1995long, toner1998flocks}. This defines a complementary {\it static} approach.   


\section{\label{dynamical method}Dynamical approach}

We begin discussing the most intuitive approach: in the absence of a driving field or any other anisotropy, collective motion is achieved by spontaneous breaking of the continuous rotational symmetry. The mean direction of any finite flock, however, is not constant, but freely diffuses since small transverse perturbations are not dumped and free to propagate (these are the so-called {\it Nambu-Goldstone} modes).\\

On the other hand, when collective motion is driven by an external field breaking rotational isotropy, fluctuations of the mean direction are confined and do not lead to free diffusion. Thus, one may discriminate between spontaneous and driven collective motion of any finite flock by simply observing the mean direction dynamics for a sufficiently long timescale. In the following, we precisely quantify this idea by developing a closed stochastic equation for the mean flocking direction.

\subsection{Mean field dynamics of the flock's orientation}

For simplicity, we work in two spatial dimensions and consider the prototypical collective motion model introduced by Vicsek and coworkers \cite{vicsek1995novel} in the presence of an homogeneous external field of amplitude $h$ and orientation $\theta_h$, ${\bf h}= h (\sin\theta_h,\cos\theta_h)$ \cite{kyriakopoulos2016leading}. It describes the discrete-time stochastic dynamics of $N$ active particles with position ${\bf r}_i^t$ and unit direction $\textbf{s}_i^t=(\cos\theta_i^t,\sin\theta_i^t)$,

\begin{equation}\label{ev_r}
{\bf r}_i^{t+1} = {\bf r}_i^t + v_{\rm 0} {\bf s}_i^t 
\end{equation}
\begin{equation}\label{micro_evo}
    \theta_i^{t+1}= \mbox{Arg}\bigg[\bigg(\sum_{j \sim i} \textbf{s}_j^t +\textbf{h}\bigg) \bigg] + \eta_i^{t},
\end{equation}
where $v_{\rm 0}$ is the particles speed and $\mbox{Arg}({\bf v})$ gives the angle defining the orientation of ${\bf v}$. Moreover, $\eta_i^t$ is a microscopic zero-average white noise such that $\langle \eta_i^t \eta_j^{t'}\rangle = \Gamma \delta_{ij}\delta_{tt'}$ and the sum is intended over the $m_j^t$ neighbours of particle $i$ (including $i$ itself) at time $t$. The neighbouring criteria may be either metric, such that $\lvert {\bf r}_i^t - {\bf r}_j^t \rvert < R_0$, or topological \cite{ginelli2010relevance}. 

We consider the Vicsek model (VM) \eqref{ev_r}-\eqref{micro_evo} in the homogeneous and highly ordered regime, deep in the so-called Toner and Tu (TT) phase \cite{ginelli2016physics}.
In the presence of an external field, the direction of motion $\theta_i^t$ of particle $i$ can be expressed in terms of deviations  $\delta \theta_i^t$ from the field direction,  $\theta_i^t = \theta_h +\delta \theta_i^t$. If we assume $\delta \theta_i^t \ll 1$, which is reasonable in the ordered phase, we can expand \eqref{micro_evo} in a {\it spin-wave} approximation \cite{nishimori2010elements}.\\
To the first order in $\delta\theta$ we get 
\begin{equation}\label{micro_evo_2}
        \theta_i^{t+1}  \approx \text{Arg} \bigg[ \hat{e}_{\parallel} \bigg(\sum_{j \sim i}1 + h \bigg)  +\hat{e}_{\perp}\sum_{j \sim i} \delta\theta_{j}^{t}  \bigg] +\eta_{i}^{t}.
\end{equation}
where $\hat{e}_{\parallel}=(\cos\theta_h,\sin\theta_h)$ is the unit vector identifying the direction of the field ${\bf h}$ and $\hat{e}_{\perp}=(-\sin\theta_h,\cos\theta_h)$ its perpendicular unit vector.\\
We note that in Eq. \eqref{micro_evo_2} the component along the perpendicular direction is small with respect to the longitudinal one. We can thus expand the Arg function as Arg($\textbf{v}+\delta \textbf{w} ) \approx $Arg$(\textbf{v}) + \frac{\textbf{v} \vee \delta \textbf{w}}{\lvert v \rvert^2}$ where $\delta {\bf w} \ll {\bf v}$ and $\vee$ denotes the \textit{skew product}\footnote{For vectors ${\bf a}, {\bf b}$ lying in a plane perpendicular to a unit vector $\hat{\bf e}_3$ one has $\textbf{a}\vee \textbf{b}=\hat{\bf e}_3 \cdot(\textbf{a}\times\textbf{b})$.}\\
By the above first order expansion Eq. \eqref{micro_evo_2} becomes
\begin{equation}\label{arg_expansion}
    \theta_i^{t+1} \approx \theta_h + \frac{\sum_{j\sim i}\delta \theta_i^t}{m_i^t+h}+\eta_i^t
\end{equation}
where $m_i^t$ is the number of neighbours of particle $i$ (including particle $i$ itself) at time $t$.
Further expanding the denominator for $h \ll m_i^t$ (we assume a sufficiently high local density, as in many systems of interest such as confluent tissues \cite{Giavazzi2018})
 \begin{equation}\label{theta_h_tilde}
     \delta\theta_i^{t+1} \approx \frac{1}{m_i^t}\sum_{j \sim i } \Big( 1-\Tilde{h}_i^t \Big)\delta\theta_j^t+\eta_i^t
 \end{equation}
where we have defined $\Tilde{h}_i^t \equiv h/m_i^t$.\\
The flocking mean direction $\psi(t)$ can be similarly expanded around the field direction 
\begin{equation}   \label{psi_t}
    \psi(t) \equiv Arg(\Omega(t)) \approx \theta_h +\frac{1}{N}\sum_{i=1}^N\delta\theta_i^t
\end{equation}
where $\Omega(t) \equiv \frac{1}{N} \sum_{i=1}^{N} \textbf{s}_i$ is the flock's global order parameter. Eq. \eqref{psi_t} implies
\begin{equation}\label{deltapsi_t}
    \delta\psi(t)\equiv\psi(t) - \theta_h \approx \frac{1}{N}\sum_{i=1}^N\delta\theta_i^t
\end{equation}

Feeding Eq. \eqref{theta_h_tilde} in \eqref{deltapsi_t} we thus obtain
\begin{equation}  \label{psi_tp1_t}
        \delta \psi(t+1) \approx \frac{1}{N} \sum_{i=1}^{N} \frac{1}{m_i^t}(1-\Tilde{h}_i^t)\sum_{j \sim i}\delta\theta_j^t +\xi(t),
\end{equation}
where we have defined $\xi(t) \equiv\frac{1}{N}\sum_{i=1}^N \eta_i^t$ and for the central limit theorem we have that
\begin{equation}
   \langle \xi(t) \rangle =0 \;,\;\;\;\;\langle \xi(t) \xi(t') \rangle = \frac{\Gamma}{N}\delta_{tt'}
\end{equation}
In order to obtain a closed equation for $\delta \psi^t$, we finally resort to a mean field approximation, $m_i^t \approx \bar{m} \equiv \langle m_i^t\rangle_{i,t}$ where the latter average is conducted over all particles $i$ and time $t$. Noting that in \eqref{psi_tp1_t} every fluctuation $\delta\theta_j$ appears roughly $\bar{m}$ times we get
\begin{equation}\label{OU_samp}
    \delta \psi(t+1) \approx  \delta\psi(t) (1-h')+ \xi(t)\,,
\end{equation}
where we have defined the reduced field amplitude $h'\equiv h/\bar{m}$.\\
Eq. \eqref{OU_samp} defines an auto-regressive process of order one, which is the time-discrete sampling of an {\it Ornstein-Uhlenbeck} process \cite{livi2017nonequilibrium}
\begin{equation}  \label{OU}
    d x(\tau) = - \gamma \, x(\tau) d\tau + \sqrt{2 D} \, d W_\tau 
\end{equation}
where $d W_\tau$ is a Wiener process, $h' = \gamma \Delta t$, $\Gamma/N = 2 D \Delta t$ and $\tau=t\,\Delta t$ with $\Delta t \ll 1$.  
The two processes share the same statistics, 
\begin{equation}  \label{FLutt_OU}
 \langle \delta \psi(t)^2 \rangle = \langle x(\tau)^2 \rangle =
 \frac{\Gamma}{2 N} \frac{1}{h'} \left(1-e^{-2 h' t}\right) \,,
\end{equation}
showing that the mean flocking direction behaves as a one dimensional Brownian particle in an harmonic potential with stiffness proportional to the external field amplitude $h$. For large times, $t \gg 1/(2 h')$, the process is stationary with
\begin{equation}  \label{FLutt_OUS}
 \langle \delta \psi(t)^2 \rangle =
 \frac{\Gamma}{2 N} \frac{1}{h'}  \,,
\end{equation}
while for $t \ll 1/(2 h')$ we recover diffusive behavior
\begin{equation}  \label{FLutt_OUD}
 \langle \delta \psi(t)^2 \rangle =
 \frac{\Gamma}{ N} \,t  \,.
\end{equation}
For $h\to 0$ therefore, our mean field approximation yields a free diffusive behavior. One may indeed repeat the above argument in the zero external field case with a spin wave expansion around the mean flocking direction, $\theta_i^t = \psi(t) + \delta \theta_i^t$. By the same mean field approximation, one finally obtains the discrete time diffusive dynamics
\begin{equation}
    \psi(t+1) = \psi(t) +  \xi(t) \,.
\end{equation}

Arguably, our mean field approximation is rather crude; in particular, by assuming a constant number of interacting neighbours, $m_i^t \approx \bar{m}$, it ignores the non-reciprocal part of Vicsek model (VM) interactions \cite{vitelli2021, chepizhko2021revisiting}. However, it should be noted that the comparison between flocking dynamics with reciprocal and non-reciprocal interactions carried on in Ref. \cite{chepizhko2021revisiting} mainly reveals significant differences at the onset of order and in confined geometries. Our theory, on the other hand, describes the behavior of the flock's mean direction (a global quantity) in the strongly ordered regime, where we expect our approximation to be harmless. In the next section, we will verify its correctness by direct numerical simulations.
\begin{figure}[tb!]	
 \includegraphics[width=0.48\textwidth]{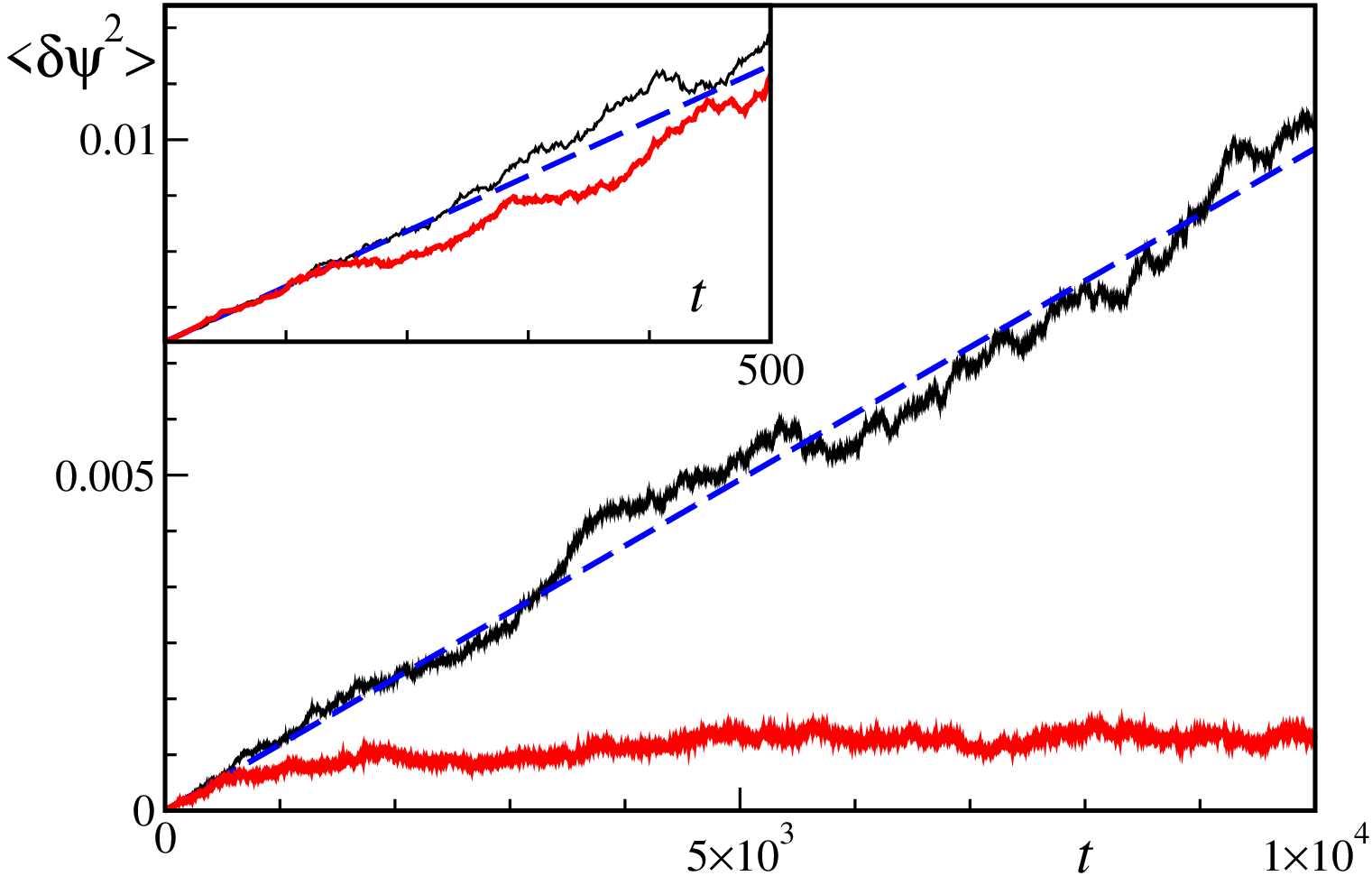}
\vspace{0.1cm}
	\caption{Spontaneous (full black line) vs. directed ($h=0.01$, full red line) mean direction squared fluctuations as a function of time. The dashed blue line is the best fit of the spontaneous symmetry breaking data (see text). We have chosen noise amplitude $\eta_{\rm 0}=0.18$ and system size $L=256$, so that $N=\rho_{\rm 0} L^2 \approx 10^5$. In the inset: zoom of the first $500$ time-steps.}
	\label{diff_spont_vs_driv}
\end{figure}
\subsection{Numerical simulations}\label{numsim}
We simulate the microscopic Vicsek dynamics \eqref{ev_r} and \eqref{micro_evo} in a two-dimensional system of linear size $L$ with periodic boundary conditions and with metric interactions. We take the white noise term $\eta_i^t$ to be uniformly distributed in the interval $[-\eta_{\rm 0} \pi, \eta_{\rm 0} \pi]$, so that its variance is
\begin{equation}\label{eq_noise}
    \langle \eta_i^t \eta_j^{t'} \rangle = \Gamma \delta_{ij}\delta_{tt'} = \frac{\eta_{\rm 0}^2 \pi^2}{3}\delta_{ij}\delta_{tt'} 
\end{equation}
In the following we fix the global particle density $\rho_{\rm 0}=N/L^2=2.0$ and particle speed $v_{\rm 0}=0.5$. Noise amplitude $\eta_{\rm 0}$ is chosen so that our the zero field system lies in the homogeneous ordered phase \cite{solon2015phase} -- the so-called Toner \& Tu (TT) phase -- comfortably far away from the ordered band regime appearing as the transition to disorder is approached \cite{gregoire2004onset, chate2008collective}.

Particles positions are initialized from a uniform distribution, while initial velocity are aligned in the external field direction (or in a given direction for $h=0$). A proper transient $T_0 \approx 10^4$ is then discarded from the dynamics to ensure convergence to the stationary ordered state. Squared fluctuations 
\begin{equation} \label{mean_fluct}
\langle \delta \psi(t)^2\rangle =\langle [\psi(t) - \psi(0)]^2 \rangle
\end{equation}
in the flock's mean direction \eqref{psi_t} are then typically evaluated averaging $10^2$ independent runs for $t\leq T=10^4$.

As shown in Fig.~\ref{diff_spont_vs_driv}, the difference between spontaneous ($h=0$, black line) and a directed collective motion ($h=0.01$, red line) is readily evident for long enough observation times. However, if one is restricted to a shorter time interval (e.g.: $10^3$ timesteps as in the inset Fig.~\ref{diff_spont_vs_driv}), discrimination between the two cases becomes problematic (especially if one is not comparing a directed with a spontaneous case but has only access to one set of data). 

A linear fit of the spontaneous case
\begin{equation}
    \langle \delta \psi^2 \rangle \approx 2D_{\rm 0}\,t
\end{equation}
returns a diffusion constant $D_{\rm 0} \approx 5.0(1) \cdot 10^{-7}$, to be compared with our mean-field prediction \eqref{FLutt_OUD}
\begin{equation}\label{mf1}
D_{\rm 0}=\frac{\Gamma}{2 N} = \frac{\eta_{\rm 0}^2 \pi^2}{6 N}
\end{equation}
\begin{figure}[tbp!]
\includegraphics[width=0.46\textwidth]{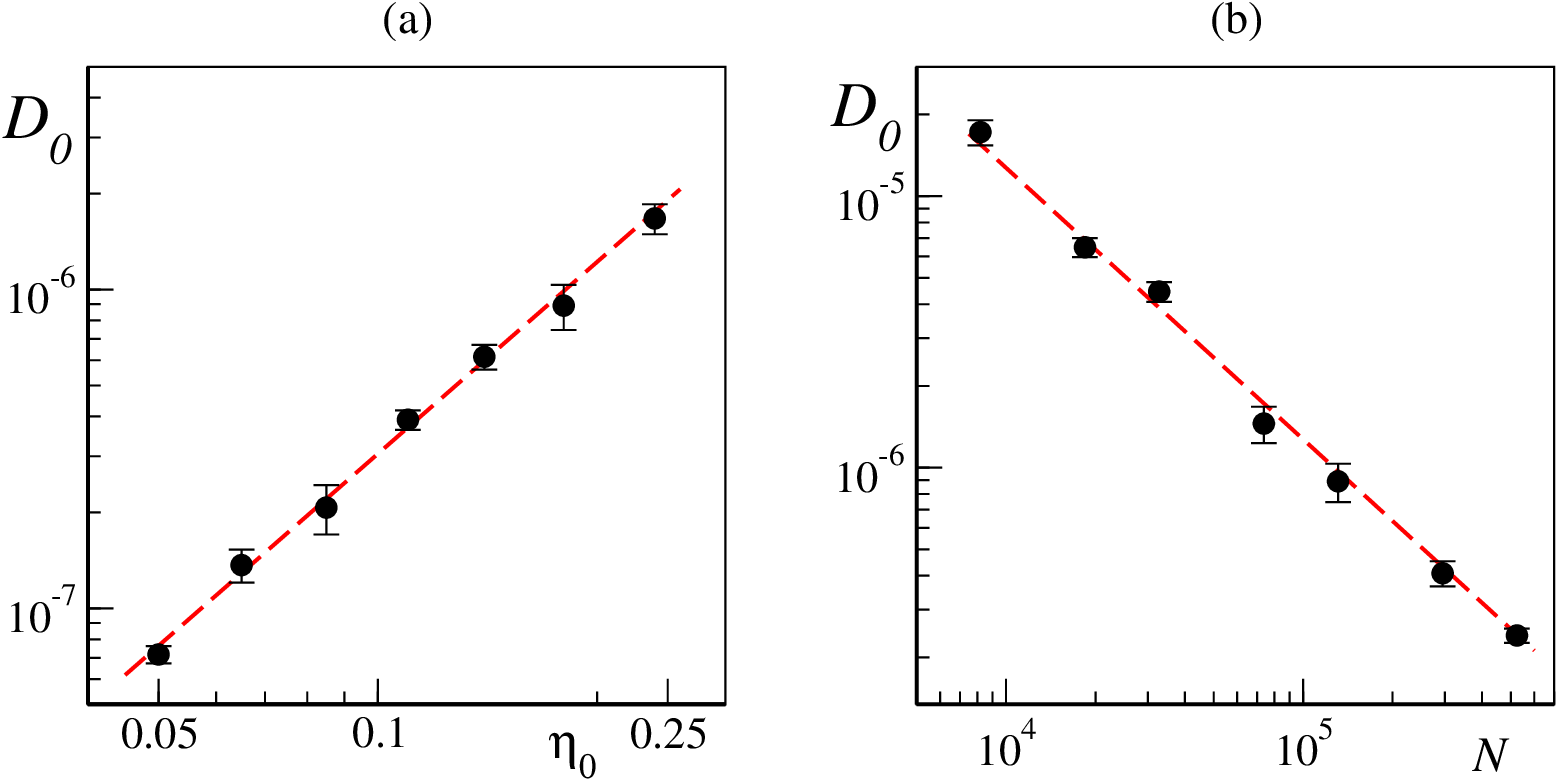}
\vspace{0.1cm}
\caption{Scaling of the diffusion constant in the zero field case. (a) Diffusion vs. noise amplitude for system size $L=256$. The dashed red line marks quadratic growth, $D_{\rm 0}\sim \eta_{\rm 0}^2$. (b) Diffusion vs. total particle number for noise amplitude $\eta_{\rm 0}=0.18$ and constant density. The dashed red line marks $1/N$ decay.
Error bars are given by two standard errors and plot are in a double logarithmic scale.} 
\label{Fig2}
\end{figure}
where we have made use of Eq. \eqref{eq_noise} and, in $d=2$, $N=\rho_{\rm 0}L^2$. With our choice of parameters, Eq. \eqref{mf1} gives $D_{\rm 0} \approx 4.1\cdot 10^{-7}$. Despite being around $20\%$ off quantitatively, Fig. \ref{Fig2} shows that the qualitative scaling with noise amplitude and particles number predicted by mean field theory is nicely verified. \\

\begin{figure}[b!]
 \includegraphics[width=0.24\textwidth]{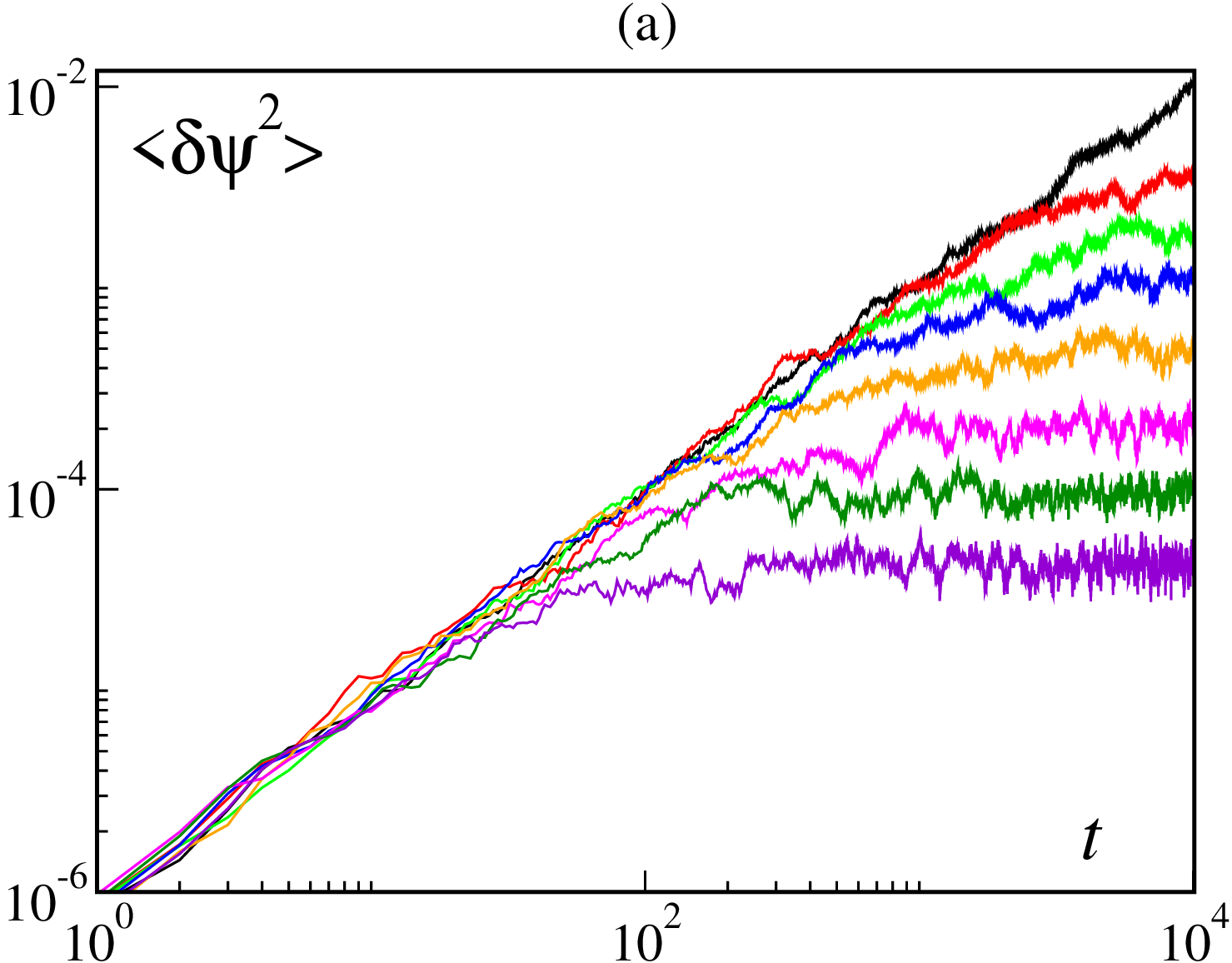}
    \includegraphics[width=0.23\textwidth]{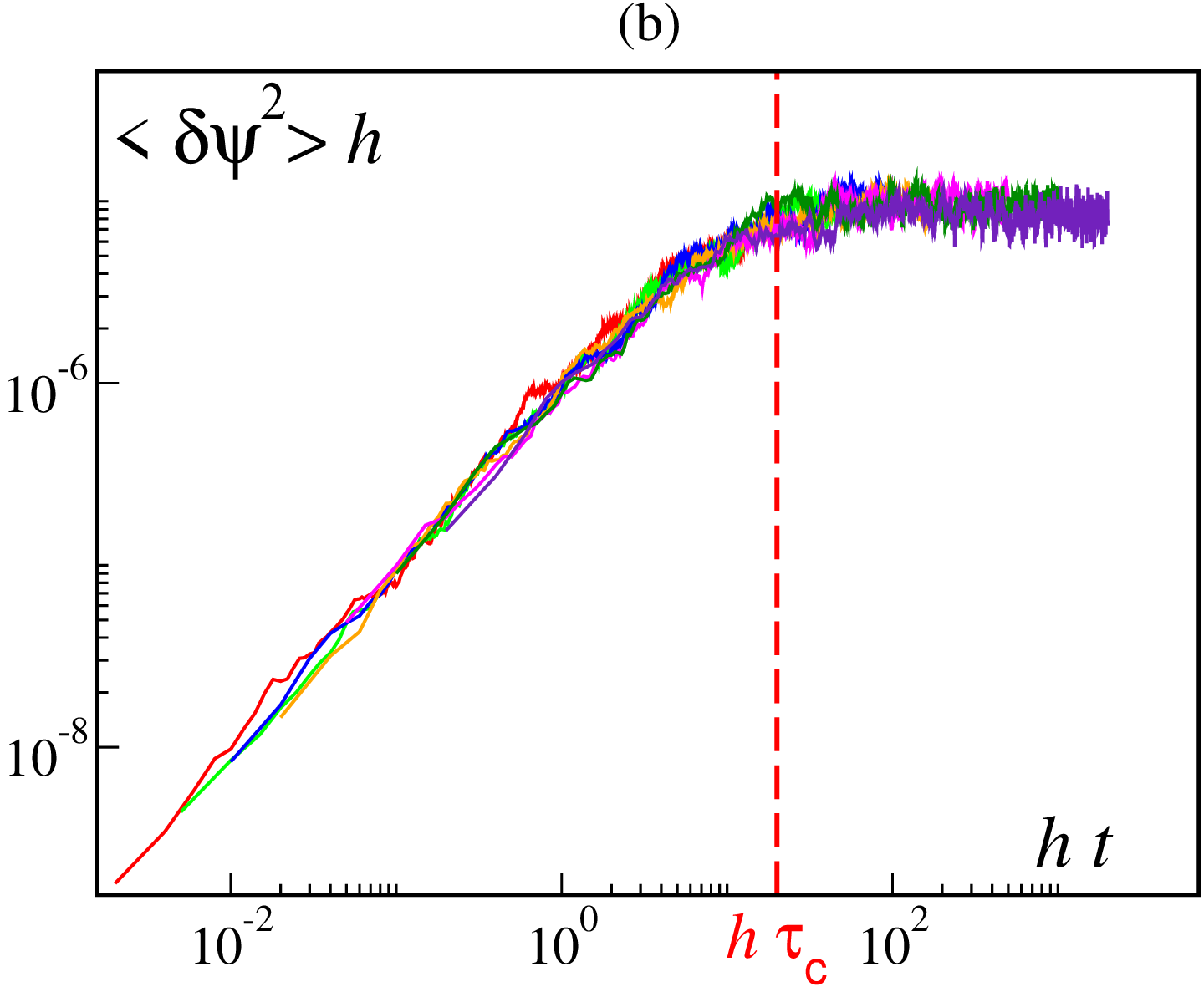}
    \caption{(a) Mean squared fluctuations of the flocking average direction as a function of time $t$ for different external field values. From bottom to top $h=0.2, 0.1, 0.05, 0.02, 0.01, 0.005, 0.002, 0.001, 0$. Data has been obtained with noise amplitude $\eta_{\rm 0}=0.18$ and system size $L=256$. 
    (b) Same data with both axes rescaled by $h$ (the $h=0$ case has been removed). The dashed red line marks the rescaled crossover time $h \tau_c$. Both graphs are in a doubly logarithmic scale.}
    \label{all_collassi}
\end{figure}

We now turn to the directed case scaling, testing different values of the field intensity $h$. Our result are shown in Fig.~\ref{all_collassi}. According to Eqs. \eqref{FLutt_OUS}, \eqref{FLutt_OUD}, rescaling by $h$ both time and the mean squared fluctuations nicely collapses the curves obtained with different $h \in [0.001,0.2]$.  This confirms that the minimum observation time needed to discriminate spontaneous from directed motion scales as the inverse of the field amplitude, with a crossover time
\begin{equation}\label{tauc}
\tau_c(h) \sim h^{-1}\,.
\end{equation}
Note that the number of active particles $N$ controls the magnitude of mean direction fluctuations but not the scaling of the crossover time.
\begin{figure}[tbp!]
\includegraphics[width=0.48\textwidth]{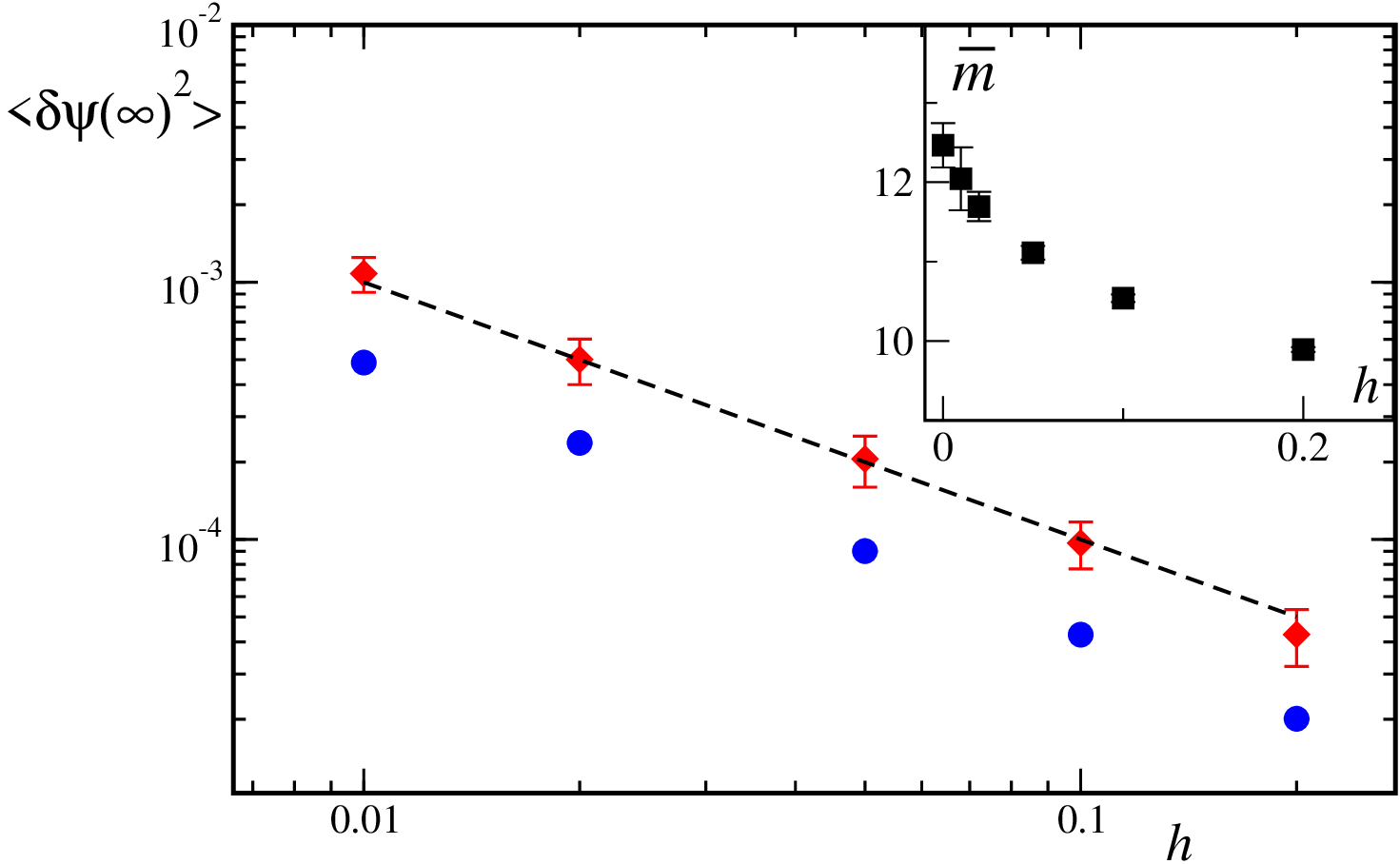}
\vspace{0.1cm} 
  	\caption{(a) Asymptotic value of $\langle\delta\psi^2\rangle$ as a function of $h$. Red diamonds report numerical estimates, while blue dots are the mean field estimate of Eq. \eqref{FLutt_OUS2} (see text). The dashed black line marks the $\sim 1/h$ scaling. Here $\eta_{\rm 0}=0.18$ and $L=256$ and the plot is in double logarithmic scale. In the inset: linear plot of the mean number of interacting active particles as a function of $h$. Error bars are given by two standard errors.}
   	\label{num_theo}
\end{figure}

Finally, we verify quantitatively the asymptotic expression for the flock's direction mean squared fluctuations \eqref{FLutt_OUS} that, by our mean field approximation reads
\begin{equation}  \label{FLutt_OUS2}
 \langle \delta \psi(\infty)^2 \rangle =
 \frac{\Gamma}{2 N} \frac{\bar{m}}{h} = \frac{\eta_{\rm 0}^2 \pi^2}{6 N} \frac{\bar{m}}{h}\,.
\end{equation}
We estimate $\bar{m}$ from direct numerical simulations and use it to compute the mean field prediction for $\langle \delta \psi(\infty)^2 \rangle$ from Eq. \eqref{FLutt_OUS2}. They are reported in Fig.~\ref{num_theo} as blue dots. The numerically measured values (red diamonds) turn out to be larger by a factor 2, possibly due to the inability of the mean field fluctuations to properly account for the anomalously large \cite{ginelli2016physics} local fluctuations in the number of interacting neighbours. Note also that numerical estimates of $\bar{m}$ (inset of Fig.~\ref{num_theo}) exhibit a weak dependence on $h$. \\

In any case, we conclude that despite the quantitative mismatch, mean field theory faithfully captures the Ornstein-Uhlenbeck scaling with $h$ (the external field playing the role of a stiffness) of the mean squared fluctuations in the flocking direction. More generally, our analysis shows that one is able to discriminate between collective and directed motion by the analysis of the mean-squared fluctuations of the average flock's direction, but only provided the observation time is larger of a crossover threshold, scaling with the inverse of the external field amplitude. As we anticipated, this might not be possible in several scenarios where the observation time is limited and/or the magnitude of spatial anisotropy is too small.

\section{\label{static method} Static approach}
We now propose a second approach, based on the observation of spatial correlations. In particular, here we focus on the static density structure factor, which is easily accessible in experimental set ups. In the absence of any external field or other anisotropies, TT theory \cite{toner1995long, toner1998flocks} predicts a long-ranged behavior for the slow fields spatial correlations, as observed in starling flocks \cite{cavagna2010scale}. Equivalently, the static structure factors, which may be obtained by Fourier transforming spatial correlation functions, exhibits a diverging behavior at small wave numbers $q$. This happens since  fluctuations transversal to the mean directions are {\it soft modes}, i.e. they are not damped and free to propagate over arbitrary distances. They are known as {\it Nambu-Goldstone} modes \cite{nishimori2010elements}. 

In the presence of an external field, on the other hand, transversal fluctuations are damped (they acquire a "mass" in the pseudo-particles language) and spatial correlations are cut-off exponentially at a finite characteristic length scale $L_c$. Correspondingly, the structure factor's divergence for $q \to 0$ is suppressed for $q < q_c \sim 1/L_c$. The cut-off length $L_c$ depends on the external field amplitude $h$ and diverges for $h \to 0$. This is well known at equilibrium since the pioneering studies of \cite{pata73, patashinskii1979fluctuation}, but as we show here it is relatively straightforward to derive the exact scaling of the cut-off length and of the static structure factor with $h$.\\

Here we focus on the static density structure factor
\begin{equation}\label{eq:sf}
    S({\bf q})= \frac{1}{N}\left\langle \sum_{n, m} e^{i {\bf q}\cdot({\bf r}_n - {\bf r}_m)}\right\rangle
\end{equation}
where ${\bf r}_n$ with $n=1,\ldots,N$ are the position of the $N$ active particles, $i$ the imaginary unit and in principle the average $\langle \cdot \rangle$ should be taken over different experimental realizations or uncorrelated snapshots of the same experiment.
In particular, its average over all wavevector orientations 
\begin{equation} \label{eq:sf_iso}
    S(q)\equiv \langle S({\bf q})\rangle _{|{\bf q}|=q}
\end{equation}
can be measured relatively easily in experimental data \cite{giavazzi2017giant}. 

Introducing the number density $\rho({\bf r},t)=\sum_n \delta ({\bf r}-{\bf r}_n^t)$ and the density fluctuations $\delta \rho({\bf r},t)=\rho({\bf r},t)-\rho_0$, the density structure factor becomes
\begin{equation}\label{eq:sf2}
S({\bf q},h) = \frac{1}{N} \langle \lvert \delta \hat{\rho}(\textbf{q},t) \rvert^2 \rangle\,.
\end{equation}
For a zero external field, Toner $\&$ Tu  theory \cite{ginelli2016physics} predicts that the isotropically-averaged density structure factor of spontaneous collective motion diverges algebraically for small wave numbers $q$ as
\begin{equation}    \label{div_S}
    S(q) \sim q^{-\zeta}
\end{equation}
where $\zeta>0$ is a universal exponent (see below). In the following, we show that, in the presence of a small and static external field of modulo  $h$, this divergence is suppressed and
\begin{equation} \label{div_S_h}
    S(q,h) \sim \frac{1}{ q^\zeta+ C \,h}
\end{equation}
with $C$ being a phenomenological constant.

\subsection{\label{theory static method}Linearized density structure factor}
We briefly recall the Toner and Tu hydrodynamic equations \cite{toner1998flocks} in the presence of a constant and homogeneous driving field ${\bf h}$ \cite{kyriakopoulos2016leading}. They rule the slow, long-wavelength dynamics of the conserved density $\rho({\bf r}, t)$ and velocity ${\bf v}({\bf r}, t)$ fields and consist in the continuity equation
\begin{equation}
\partial_t\rho +\nabla\cdot(\rho{\bf v})=0
\label{rho1}
\end{equation}
and the velocity field dynamics 
\begin{equation} \label{second_v}
    \begin{split}
        \partial_t \textbf{v} &+\lambda_1(\textbf{v} \cdot \nabla)\textbf{v} +\lambda_2 (\nabla \cdot \textbf{v})\textbf{v} + \lambda_3\nabla \lvert \textbf{v} \rvert^2 \\ =& (\alpha - \beta \lvert \textbf{v} \rvert^2 ) {\bf v} -\nabla P_1 -\textbf{v}(\textbf{v} \cdot \nabla P_2) + D_1 \nabla(\nabla \cdot \textbf{v})\\& +D_3 \nabla^2 \textbf{v} +D_2(\textbf{v} \cdot \nabla)^2 \textbf{v} + \textbf{f} +\textbf{h}\,.
    \end{split}
\end{equation}
Here all the phenomenological convective ($\lambda_i$, with $i=1,2,3$) and viscous ($D_i>0$) coefficients, as well as the two symmetry breaking ones, $\alpha$ and $\beta$ can, in principle, depend on $\rho$ and
$|{\bf v}|$ and the pressures $P_{1,2}$ may be expressed as a series in the density \cite{toner2012reanalysis}. The additive noise term ${\bf f}$ has zero mean, variance $\Delta$ and is delta correlated in space and time. The coarse-grained constant field ${\bf h}=h \hat{e}_{\parallel}$ is, by analyticity and rotational invariance of the free system, linearly proportional to the applied microscopic field, as long as those fields are sufficiently small.

In the absence of fluctuations Eqs. \eqref{rho1}-\eqref{second_v} admit a homogeneous steady state solution
$\rho({\bf r}, t)=\rho_0$,
${\bf v}({\bf r}, t)=v_{\rm 0}(h) \hat{e}_{\parallel}$\,,
where $v_0(h)$ is determined by the condition
\begin{equation}
    \alpha v_0 - \beta v_0^3 +h = 0 \,.
\end{equation}
In the zero external field case (and for $\alpha>0$), $v_0 = \sqrt{\alpha/\beta}$, with the direction $\hat{e}_{\parallel}$ randomly selected by the spontaneous symmetry breaking mechanism. Assuming analyticity of the symmetry breaking coefficients, we have for small $h$
\begin{equation}
v_0(h)-v_0(0) \propto h
\end{equation}

To deal with fluctuating hydrodynamics, we follow \cite{toner1998flocks}-\cite{toner2012reanalysis}-\cite{kyriakopoulos2016leading} and proceed to linearize Eqs. \eqref{rho1}-\eqref{second_v} around the homogeneous solution,
\begin{equation} \label{fluct}
    \begin{split}
        & \rho(\textbf{r},t)=\rho_0 +\delta \rho(h,\textbf{r}, t)\\
        & \textbf{v}(\textbf{r},t)=[{v}_{\rm 0}(h)+ \delta v_{\parallel}(h,\textbf{r},t)] \hat{e}_{\parallel}+ \textbf{v}_{\perp}(h,\textbf{r},t)
    \end{split}
\end{equation}

where $\textbf{v}_{\perp}$ measures velocity fluctuations transversal to $\hat{e}_{\parallel}$. In the Toner and Tu phase, the longitudinal velocity fluctuations $\delta v_\parallel$ are a fast mode enslaved to the slow fields of the unperturbed theory, $\textbf{v}_{\perp}$ and the density fluctuations $\delta \rho$. Thus, they can be eliminated from Eqs. \eqref{rho1}-\eqref{second_v} to yield the linearized hydrodynamics
\begin{equation}\label{rho_lin}
\begin{split}
        \partial_t \delta\rho &+\rho_0 \nabla_{\perp}\cdot \textbf{v}_{\perp} +v_2 \partial_{\parallel}\delta\rho -\rho_0 \mu_2\partial_t \partial_{\parallel}\delta\rho \\
        & =D_{\rho_{\parallel}} \partial_{\parallel}^2 \delta\rho + D_{\rho_{\perp}} \nabla_{\perp}^2 \delta\rho +D_{\rho v}\partial_{\parallel} (\nabla_{\perp} \cdot \textbf{v}_{\perp})  \\ 
\end{split}
\end{equation}
and
\begin{equation}  \label{v_perp_lin}
    \begin{split}
        \partial_t \textbf{v}_{\perp} &+ \gamma \partial_{\parallel} \textbf{v}_{\perp}\\ & = -\frac{c_0^2}{\rho_0}\nabla_{\perp} \delta\rho +D_B \nabla_{\perp}(\nabla_{\perp}\cdot \textbf{v}_{\perp}) +D_3 \nabla_{\perp}^2 \textbf{v}_{\perp} \\ & +D_{\parallel}\partial_{\parallel}^2\textbf{v}_{\perp} 
        +g_t \partial_t \nabla_{\perp}\delta\rho + g_{\parallel} \partial_{\parallel} \nabla_{\perp}\delta\rho +\textbf{f}_{\perp}-h_v \textbf{v}_{\perp} 
    \end{split}\,,
\end{equation}
where we have introduced the reduced field
\begin{equation}\label{reduced-f}
    h_v \equiv \frac{h}{v_{\rm 0}(0)}
\end{equation}
and all the various constants introduced above can be expressed as a function of the phenomenological constants appearing in Eqs. \eqref{rho1}-\eqref{second_v}. Their precise expressions are not important in what follows, but they can be found -- together with all the details of the linearization -- in Ref. \cite{kyriakopoulos2016leading}.

The linearized density structure factor can straightforwardly computed from \eqref{rho_lin} and \eqref{v_perp_lin} in Fourier space, where
\begin{equation}\label{fourier_campi}
    \begin{split}
        &\hat{\bf v}_\perp(\textbf{q},\omega)\sim \int d\textbf{r}dt \,e^{-i(\textbf{q}\cdot \textbf{x}-\omega t)}\textbf{v}_\perp(\textbf{r},t)\,,\\ 
        & \delta\hat{\rho}(\textbf{q},\omega) \sim \int d\textbf{r}dt\, e^{-i(\textbf{q}\cdot \textbf{x}-\omega t)}\delta \rho(\textbf{r}, t)\,.
    \end{split}
\end{equation}

This calculation closely resembles the one for the zero field case \cite{toner2012reanalysis}, and for compactness we report its details in the appendix. To leading order in wave vector ${\bf q}$ the linearized structure factor is
\begin{equation} \label{autocorr_eqt}
\begin{split}
        S({\bf q},h)\equiv \langle \lvert \delta \hat{\rho}(\textbf{q},t) \rvert^2 \rangle=&\frac{1}{2} \bigg( \frac{\Delta\, \rho_0^2 \,\sin(\theta_q)^2}{[c_+(\theta_q)-c_-(\theta_q)]^2} \bigg) \\
        &\times \bigg( \frac{1}{\epsilon_+(h,{\bf q})}+ \frac{1}{\epsilon_-(h,{\bf q})} \bigg)\,,
        \end{split}
\end{equation}
where  $q=|{\bf q}|$ and $\theta_q$ is the angle between $\textbf{q}$ and $\hat{e}_\parallel$. We have also introduced the sound speeds $c_\pm(\theta_q)$ and the field dependent dampings
\begin{equation}\label{epsilon_pm}
\epsilon_\pm (h, {\bf q}) = \epsilon_\pm (0, {\bf q})+ a_\pm(\theta_q) h \equiv b_\pm(\theta_q) q^2 + a_\pm(\theta_q) h\,,
\end{equation}
where
\begin{equation}\label{eq:apm}
a_\pm(\theta_q)=\frac{(c_{\pm}(\theta_q)-v_2\cos(\theta_q)^2)}{v_{\rm 0}(0)[2c_{\pm}(\theta_q)-(v_2+\gamma)\cos(\theta_q)]}\,.
\end{equation}
They are respectively the real ($c_\pm$) and imaginary part ($\epsilon_\pm$) of the linearized system eigenfrequencies (see appendix).
The precise form of $c_\pm(\theta_q)$ and $b_\pm(\theta_q)$ is not relevant for what follows and its reported in the appendix for compactness. Here it suffice to note that they are only a function of the angle $\theta_q$.\\
We conclude that a small static and homogeneous external field only affects the damping terms, which acquire a correction linear in the field amplitude but independent of the wavevector magnitude $q$. This suppresses the small wavelength divergence of the free theory, since
\begin{equation} \label{autocorr_q0}
       \lim_{q \to 0} S({\bf q},h) \sim h^{-1}
\end{equation}

\subsection{Nonlinear corrections}
The structure factor \eqref{autocorr_eqt} is only valid in linear approximation. Nonlinear terms, ignored in the linearized approach, are known to be relevant in $d<d_c=4$ \cite{toner1998flocks,toner2005hydrodynamics} and can be accounted for by a dynamical renormalization group (DRG) analysis \cite{forster1977large, DRG2}. The original DRG analysis of field-free flocks has been carried on in \cite{toner1998flocks}-\cite{toner2012reanalysis}, while the driven case has been first analyzed in \cite{kyriakopoulos2016leading} in order to compute linear response. Here we follow the same approach to compute the scaling behavior of the density structure factor.\\

The DRG procedure is carried on in the transversal directions and consist in two steps. In the first one, the nonlinear equations of motion are averaged over the short-wavelength fluctuations: i.e., we average over the slow fields Fourier modes \eqref{fourier_campi} with wavevector lying in the (hyper-cylindrical) shell of Fourier space $b^{-1} \Lambda \!
\le q_{_\perp}  \le \Lambda$. Here
$\Lambda$ is an ``ultra-violet cutoff'' (essentially dictated by the inverse of the microscopic interaction range), and $b>1$ is an arbitrary rescaling factor. Here and in the following we use the subscripts $\perp$ and $\parallel$ to denote, respectively, the directions perpendicular and parallel to the broken symmetry one, i.e. $q_\perp=|{\bf q}_\perp|=q \sin(\theta_q)$ and $q_\parallel=q \cos(\theta_q)$.\\

In the second step, in order to restore the ultraviolet cut-off to $\Lambda$, one rescales transversal distances $r_\perp=|{\bf r}_\perp |$ and wave numbers as 
\begin{equation}
r_\perp = b r_\perp'\;\;\;\;,\;\;\; q_\perp = b^{-1}q_\perp'\,.
\end{equation}
Also time, parallel distances and the fields rescale according to \footnote{One may show that transversal velocity and density fluctuations have the same scaling \cite{toner1998flocks}}
\begin{equation}
r_\parallel = b^{\xi} r_\parallel'\;,\;\;\; q_\parallel = b^{-\xi} q_\parallel'\;,\;\;\;
t= b^{z}t'\;,\;\;\;\delta \rho = b^{\chi} \delta \rho'\,.
\end{equation}
This procedure leads to a new "renormalized" set of equations of motion with the same form w.r.t. to the original ones but with 'renormalized' parameter values. If we denote collectively the initial full parameter set of the nonlinear equations of motion as $\{\mu_i^{(1)}\}$, we can represent their evolution by the above DRG flow as $\{\mu_i^{(1)}\}\to \{\mu_i^{(b)}\}$. The scaling exponents  $\xi$, $z$, and $\chi$, known respectively as the "anisotropy", "dynamical" and "roughness" exponents, are in principle arbitrary. For a suitable choice of their value, however, a "renormalization group fixed point" 
$\{\mu_i^*\}$, that is, a situation in which the renormalized parameters do not change under this renormalization group process, is obtained in the $b \to \infty$ limit. Analyzing the DRG flow near this fixed point one can thus deduce the system scaling properties.\\

The averaging step has to be performed perturbatively in the equations' of motion nonlinearities, and generally produces nonlinear corrections (the so-called {\it graphical} corrections) in the DRG flow equations. This makes arduous an exact treatment of the nonlinear DRG fixed point, and indeed no exact values for the zero external field Toner \& Tu theory scaling exponents at the nonlinear fixed point are known below the upper critical dimension \cite{toner2012reanalysis}. Recent high precision simulations of the microscopic Vicsek model, however, provide fairly good estimates in $d=2,3$ \cite{mahault2019quantitative}. 

Including an external field in the DRG analysis is, as usual, fairly simple, and we actually may discuss it without the need to specify the exact form of the nonlinear equations of motion (they may be checked in \cite{kyriakopoulos2016leading}). It suffice to know that they have to be rotationally invariant with the only exception of the rescaled field $h_v$, the only term explicitly breaking the rotational symmetry. As a consequence, it may not gain any graphical correction in the short wavelength averaging step. Moreover, trivial dimensional analysis of Eq. \eqref{v_perp_lin} shows that $h_v$ scales as the inverse of time, yielding the linear recursion
\begin{equation}
    h_v \to h_v'=b^{z} h_v
\end{equation}
We are now able to write down the recursive equation for the density structure factor in the presence of an external field,
\begin{equation}\label{scaling1}
S({\bf q}_\perp, q_\parallel, \{\mu_i^{(1)}\}, h_v) = b^{\zeta} S(b {\bf q}_\perp, b^{\xi} q_\parallel, \{\mu_i^{(b)}\}, b^{z} h_v)\,,
\end{equation}
where we have introduced $\zeta\equiv(d-1) + \xi + 2 \chi$ and the scaling of the structure factor has been determined considering that it is given by the Fourier transform of the equal time, real space density correlation function. Thus, it involves two powers of the density fluctuations and one volume element. 

The scaling of the structure factor with the field can be now deduced by fixing a reference value $h_v^*$ for the reduced field \eqref{reduced-f} and choosing the rescaling factor $b$ such that $b^{z} h_v=h_v^*$, which
implies 
\begin{equation}
    b=\left(\frac{h_v}{h_v^*}\right)^{-1/z}\,.
\end{equation}
In practice, one adapts the DRG magnifying glass to the value of the external field \cite{nishimori2010elements}. Furthermore, if $h\ll 1$ and thus also $h_v$, this choice implies $b\gg 1$, so that all other parameters $\{\mu_i\}$ will flow to their fixed point value, $\{\mu_i^{(b)}\} \to \{\mu_i^*\} $. Therefore, from Eq. \eqref{scaling1} we have 
\begin{equation}\label{scaling2}
S({\bf q}_\perp, q_\parallel, \{\mu_i^{(1)}\}, h) = h^{-\zeta/z} g(h^{-1/z} {\bf q}_\perp, h^{-\xi/z} q_\parallel)\,,
\end{equation}
where we have introduced the universal scaling function
\begin{equation}\label{scaling-g}
g({\bf x},y)=b^{\zeta/z} S(h_{\rm 0}^{1/z} {\bf x}, h_{\rm 0}^{\xi/z} y , \{\mu_i^*\}, h_v^*) \,,
\end{equation}
and $h_{\rm 0}=h_v^* v_{\rm 0}(0)$.

While for an anisotropy exponent $0<\xi<1$ correlations behave differently in the longitudinal and transversal directions, it may be convenient to consider the isotropically-averaged density structure factor 
\begin{equation}
    S(q,h) = \langle S({\bf q},h)\rangle_{|{\bf q}|=q}
\end{equation}
which is more easily accessible in experimental measures \cite{giavazzi2017giant}. Note also that the most accurate numerical estimates of correlation functions suggest little or no anisotropy in $d = 2,3$ dimensions \cite{mahault2019quantitative}. In any case, Toner and Tu theory predicts that the short wavelength structure factor is dominated by contributions in the $q \sim q_\perp \sim q_\parallel$ direction \cite{ginelli2016physics} so that
\begin{equation}\label{s_free}
    S(q,0) \sim q^{-\zeta}\,.
\end{equation}
Along the $q_\perp \sim q_\parallel$ line and for small $h$ one has $h^{-1/z} q \sim h^{-1/z} q_\perp \gg h^{-\xi/z} q_\parallel$, so that the structure factor is dominated by transverse wavevectors ${\bf q}_\perp$ and we can replace $g$ with the universal scaling function $w(x)=g(x,0)$ (if $\xi=1$ one otherwise chooses $w(x)=g(x,x)$). This finally yields the scaling for the isotropic structure factor
\begin{equation}\label{scaling3}
S(q, h) = h^{-\zeta/z} w(h^{-1/z} q)\,.
\end{equation}

The behavior of the universal scaling function $w$ can be inferred by the request that, for $h\to0$, the structure factor's scaling coincides with the one of Eq. \eqref{s_free}, so that
\begin{equation}
w(x) \sim \left\{ \begin{array}{cc}
   x^{-\zeta}&\;\;\;\mbox{for}  \;\;x\gg1    \\
 \mbox{const.} & \;\;\;\mbox{for}\;\; x \ll 1
\end{array}\right.\,, 
\end{equation}
and the isotropically averaged structure factor takes the form
\begin{equation} \label{scaling5}
    S(q,h)\sim  \frac{h^{-\zeta/z}}{h^{-\zeta/z} q^\zeta +C} = \frac{1}{q^\zeta+C\, h^{\zeta/z}}\,.
\end{equation}
where $C$ is a phenomenological parameter.
Numerical estimates of the scaling exponents \cite{mahault2019quantitative} give
\begin{equation}\label{hypers}
z=\zeta = \left\{ \begin{array}{cc}
   1.33(2) &\;\;\;\mbox{in}  \;\;d=2    \\
 1.77(3) & \;\;\;\mbox{in}\;\; d=3
\end{array}\right.\,, 
\end{equation}
which strongly supports the hyperscaling relation 
\begin{equation}\label{hyper}
z=d-1+\xi+2\chi=\zeta
\end{equation}
conjectured in \cite{toner1995long, toner1998flocks, toner2012reanalysis}. If this is the case, Eq. \eqref{scaling5} simplifies to finally yield Eq. \eqref{div_S_h}.

\begin{figure}	
	\includegraphics[width=0.45\textwidth]{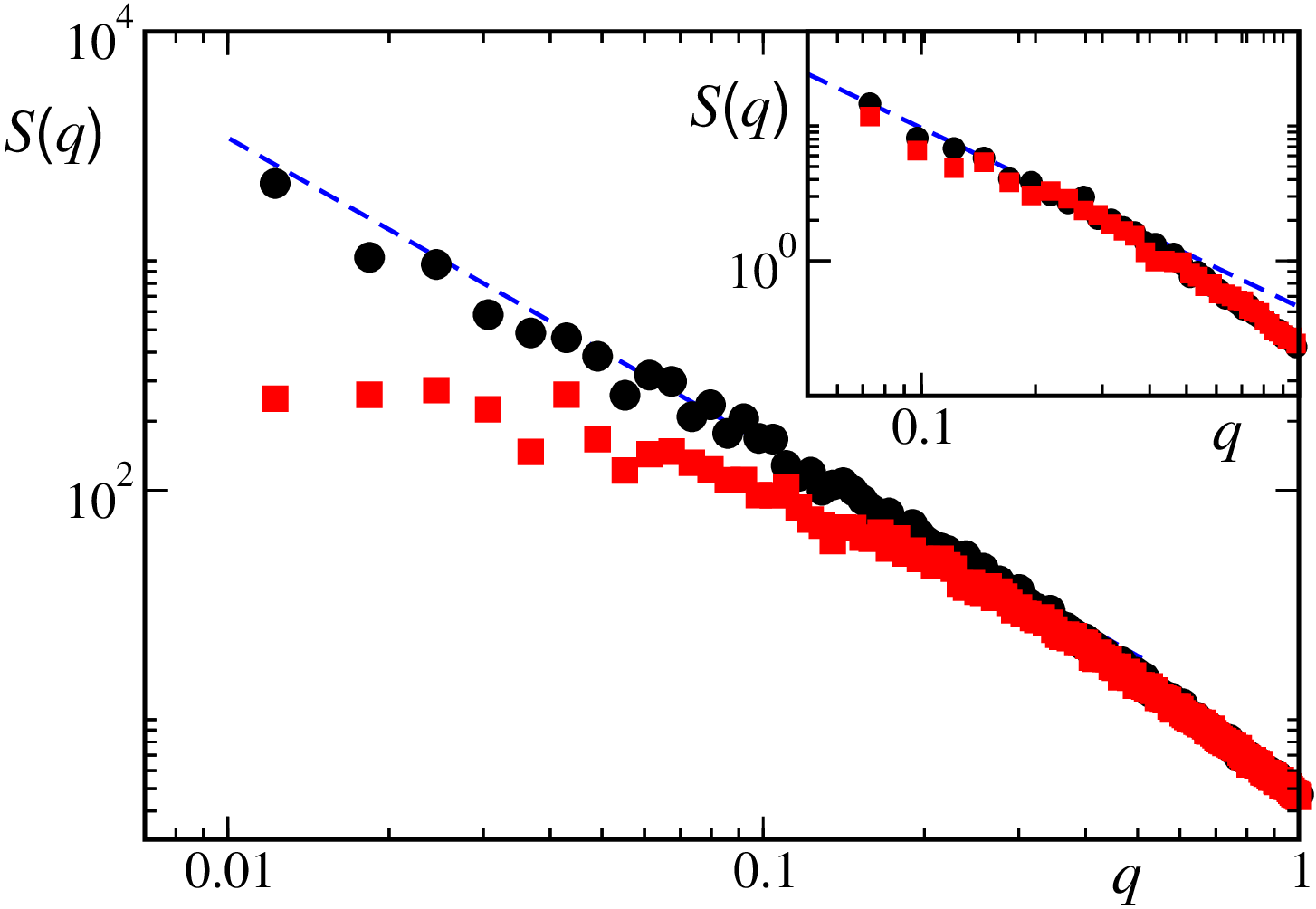}
\vspace{0.2cm}
	\caption{Isotropically averaged density structure factor $S(q,h)$ for the spontaneous (black dots) and directed ($h=0.01$, red squares) case for systems of size $L=1024$.  Data has been obtained averaging the structure factor of ten different configurations sampled every 50 timesteps (see text). The dashed blue line marks the power law divergence $\sim q^{-\zeta}$, with $z=1.33$. Other parameters are $\rho = 2.0$, $v_0=0.5$, $\eta=0.18$. Inset: same parameters as in the main panel, but with system size $L=256$. All axes are in a double logarithmic scale.}
	\label{2048_final_sq}
\end{figure}

\subsection{\label{simulations static method}Numerical scaling }

We tested our static method on two dimensional synthetic data generated from the Vicsek model \eqref{ev_r}-\eqref{micro_evo} with metric interactions. We consider systems in the stationary TT phase (in the following: $\rho=2.0$, $\eta=0.18$, $v_{\rm 0}=0.5$ with periodic boundary conditions) with or without an external field. We compute the structure factor from Eq. \eqref{eq:sf2}, starting from real space density field, coarse-grained in boxes of size one. The resulting structure factor $S({\bf q})$ is further averaged over all the ${\bf q}$ orientations to obtain its isotropic average, $S(q)$ (see Eq. \eqref{eq:sf_iso})\footnote{The isotropic average is finally binned over channels of width $2 \pi/L$.}. Invoking the ergodicity of the stochastic process, the average over different realizations may be replaced by time-averages. 

We first consider an external field of magnitude $h=0.01$. In Section \ref{numsim}, we have seen that its presence may be inferred from the dynamic of the flock's mean direction only when observation times are larger than $\tau_c \approx 500$ (see Fig. \ref{diff_spont_vs_driv}). In order to test the static approach in a regime where the dynamic one fails, we have restricted the time averages of the structure factor over a time-window of $T=500$ timesteps. Due to temporal correlations between subsequent spatial configurations, this may yield a low statistics, especially for the lowest $q$ modes, so we have cut-off frequencies $q\leq \pi/L$.  

In Fig.~\ref{2048_final_sq} we consider a relatively large ($L=1024$) system and compare results for the directed ($h=0.01$) collective dynamics with the ones for spontaneous collective motion ($h=0$). The latter case shows a behavior compatible with the free scaling $S(q) \sim q^{-\zeta}$ with $\zeta=1.33$ \cite{mahault2019quantitative}, while in the former case, the suppression of the low $q$ divergence is rather evident for $q<q_c$, where $q_c$ is a crossover wavenumber.  This shows the viability of our static approach for large enough systems.
In smaller systems ($L=256$, inset of Fig.~\ref{2048_final_sq}), on the other hand, directed collective motion cannot be detected by the observation of the low $q$ behavior since the divergence suppression becomes effective at wavenumbers not accessible due to the limited system size.

\begin{figure}[tp!]
    \includegraphics[width=0.49\textwidth]{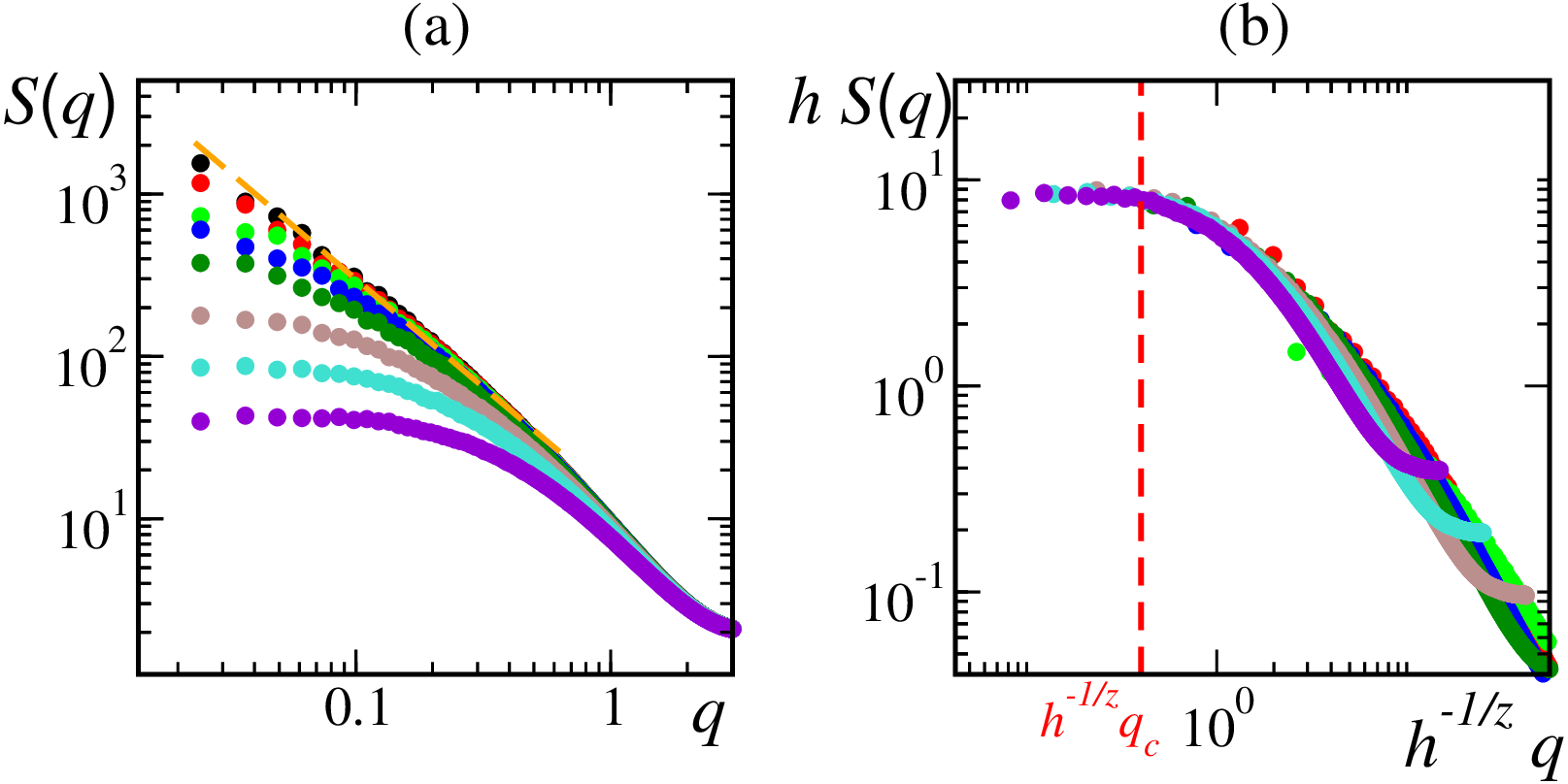}
\vspace{0.1cm}
    \caption{(a) Isotropically averaged structure factor $S(q,h)$ for different external field amplitudes $h$. From top to bottom $h=0, 0.002, 0.005, 0.01, 0.02, 0.05, 0.1, 0.2$. The dashed The dashed orange line marks the power law divergence $\sim q^{-\zeta}$, with $z=1.33$. 
    (b) same as (a), but with the data rescaled by $h^{-1/z}$ (horizontal axis) and $h$ (vertical axis) in order to collapse the structure factor curves. The $h=0$ data has been omitted.
    The vertical dashed red line marks the location of the (rescaled) crossover wavenumber (see text).
    All axes are in a double logarithmic scale, and model parameters
	are $\rho = 2.0$, $v_0=0.5$, $\eta=0.18$ and $L=512$.}
	\label{final_sq}
\end{figure}

To better probe the structure factor scaling, we now run longer time-averages for systems of size $L=512$ and different external field amplitudes $h\in [0, 0.2]$. They are shown in Fig.~\ref{final_sq}a. According to the scaling law \eqref{scaling3} and the hyperscaling relation \eqref{hypers}, the field rescaled structure factor $h S(q)$ should be only a function of $h^{-1/z} q$, and this is verified nicely in Fig.~\ref{final_sq}b where we have once again used used the 2d numerical estimate $z=1.33$. This also implies the crossover wavenumber scaling
\begin{equation}\label{crossover_size1}
    q_c(h) \sim h^{1/z}\,.
\end{equation}
Equation \eqref{crossover_size1} immediately implies that observations need to be carried on in large enough systems, with a threshold linear system size
\begin{equation}\label{crossover_size2}
    L_c(h) \sim h^{-1/z}\,.
\end{equation}
Note that the crossover temporal and spatial scales are indeed related through the dynamical exponent,
\begin{equation}\label{crossover_size3}
    \tau_c \sim L_c^z \,.
\end{equation}


\section{Conclusions}

We have discussed the behavior and scaling properties of the mean flocking direction and static density correlations in the presence of a small homogeneous external field. For large enough times, fluctuations in the mean direction depart from the diffusive behavior exhibited by finite flocks in the zero field case, following the statistics of an Ornstein-Uhlenbeck process. For the first time, we have also discussed explicitly the scaling of the diffusive process exhibited by the mean direction of motion of finite flocks in the spontaneous, zero-field case.

In large enough systems, a complementary signature of directed motion can be found in the small $q$ behavior of the static density structure factor, which saturates in the presence of an external field rather than showing the divergent behavior typical of spontaneous symmetry breaking.

These facts can be used to detect directed collective motion, that is flocking behavior guided by external cues such as concentration gradients or other global anisotropies. Observations of the mean flock's direction (the {\it dynamical} method) may be useful for long enough observation times, while computing the density structure factor (the {\it static} method) is informative in sufficiently large systems. 
Note also that in order to apply our analysis it is not necessary that all active particles are able to detect the environmental cues. In Ref \cite{giomi}, it has been shown that the effect of an external static field only affecting a finite fraction of the flocking particles is equivalent to that of a rescaled (by the fraction of affected particles) homogeneous field. The equivalence holds provided the affected particles are randomly distributed in the flock. On the contrary, we do not expect our considerations to be directly applicable to localized or time-varying perturbations \cite{cavagna2013boundary}.

It is also fair to stress that while our method can detect the presence of external fields/environmental cues, it cannot exclude their presence: negative results may simply mean that the external field is so small that the accessible spatial or temporal observational scales are too small to detect it. 
On the other hand, strong external fields or anisotropies beyond the linear regime should result in a complete suppression of the static structure factor low $q$ divergence and of the diffusive, short time behavior of the mean flock direction. In this regime, however, our scaling relations are no longer valid.

This work focused on the explicit symmetry breaking of the continuous rotational symmetry by a {\it vectorial} external field, which we believe to be the most biologically relevant situation. However, at least in principle, one may also conceive spatial anisotropies inducing more complex discrete symmetries. A prominent example is the active Ising model (AIM) \cite{Solon2013}, where active particles move on a square lattice and bias their movement along, say, the vertical axis, favouring upwards or downward hopping according to their binary spin variable. The discrete $\mathbb{Z}_2$ symmetry completely suppresses the structure factor low $q$ divergence, and strongly pins the mean flock direction either in the upward or downward direction, similar to what is expected in the presence of a strong external field. Perhaps more intriguing, is the generalisation to the so-called active clock model recently carried on in \cite{Solon2022}. Here, each particle orientation ${\bf s}_i$ can be in $Q$ different states equally distributed around the unit circle (the AIM being recovered for $Q=2$), and the hopping bias along the lattice is proportional to the projection of ${\bf s}$ in the hopping direction. The clock model is characterized by a crossover scale which grows {\it exponentially} with $Q$. Below such a crossover scale, the system shows a wandering order parameter and a diverging structure factor, while above it the order parameter is pinned to a discrete clock direction and the divergence is suppressed at low wave-numbers. This analysis is clearly complementary to ours: we have developed scaling relations for continuos particle orientations slightly biased by a small vectorial field, while Ref. \cite{Solon2022} strongly constrains orientations along a discrete set of $Q$ directions. We derived scaling w.r.t. the field intensity, while Ref. \cite{Solon2022} studies scaling w.r.t. the number of strongly constrained directions.

In any case, its is fair to note that the approaches put forward in the present paper cannot discriminate, from the analysis of a single given collective motion instance,  between collective motion in the presence of a small external field and collective motion that somehow follows the discrete symmetry of, say, a large $Q$ active clock model. Or to disentangle the role of the two effects, when both are present.
However, while our analysis cannot technically exclude the presence of more than one preferred direction of motion, we believe that this latter possibility should not be relevant in many situations of practical interest, and thus that our approach conveys useful information on the nature of collective motion.

In the future, we plan to test our method on data obtained from {\it in vitro} cellular migrations\cite{malinverno2017endocytic}  taking place on a grooved substrate, but we also expect our considerations to be useful to biologist to detect clear signatures of the directed nature collective motion in {\it in vivo} cellular migration phenomena.   

\begin{acknowledgments}
We acknowledge support from PRIN 2020PFCXPE. FG thanks Clement Zankoc for earlier numerical tests.
\end{acknowledgments}

\appendix*
\section{Linearized structure factor: technical details}

In the following we derive the linearized structure factor \eqref{autocorr_eqt} from Eqs.~\eqref{rho_lin}-\eqref{v_perp_lin}.
First, we rewrite Eqs. \eqref{rho_lin}-\eqref{v_perp_lin} in  Fourier space, according to \eqref{fourier_campi} 
\begin{equation}   \label{rho_four_first}
    \begin{split}
         &[-i(\omega-v_2 q_{\parallel}) +D_{\rho_{\parallel}} q_{\parallel}^2 +D_{\rho_{\perp}} q_{\perp}^2-\phi q_{\parallel} \omega] \delta\hat{\rho} \,+\\
         & +[i \rho_0 q_{\perp} +D_{\rho v}q_{\perp} q_{\parallel}]\hat{v}_L=0,
    \end{split}
\end{equation}

\begin{equation}\label{v_L_first}
\begin{split}
        &\Big[ \frac{i c_o^2}{\rho_0}q_{\perp} -g_t q_{\perp}\omega +g_{\parallel} q_{\perp} q_{\parallel} \Big] \delta\hat{\rho}\,+\\
        &+[-i(\omega-\gamma q_{\parallel})+ D_L q_{\perp}^2 + D_{\parallel} q_{\parallel}^2 +h_v ]\hat{v}_L = \hat{f}_L
\end{split}
\end{equation}
where we have defined $\phi \equiv\rho_0 \mu_2$, $\gamma=\lambda_1 v_0(0)$ and $D_L \equiv D_B+D_T$. 
Moreover, $\hat{v}_L$ e $\hat{f}_L$ are the components along the longitudinal direction of $\hat{\bf v}_{\perp}$ and $\hat{\bf f}_{\perp}$ (the Fourier transformed transversal noise) defined as
\begin{equation}\label{A2}
    \hat{v}_L \equiv \hat{\bf v}_\perp \cdot \frac{\textbf{q}_{\perp}}{q_{\perp}} \ \ \text{and} \ \  \hat{f}_L \equiv \hat{\bf f}_\perp \cdot \frac{\textbf{q}_{\perp}}{q_{\perp}}.
\end{equation}
Notice that we have  omitted the equation for the $(d-2)$ transversal modes $\hat{\bf v}_T$, which are the components of $\hat{\bf v}_\perp$ orthogonal to ${\bf q_\perp}$: this is simply due to the fact that it is decoupled from Eqs. \eqref{rho_four_first}-\eqref{v_L_first} and it does not contribute to the longitudinal eigenmodes and the long-ranged behavior of density correlations \cite{toner2012reanalysis}.

We proceed to find the normal modes eigenfrequencies $\omega(\textbf{q})$ of Eqs.\eqref{rho_four_first}-\eqref{v_L_first}, that is, the complex frequencies at which non-zero solutions exist for zero noise, $\hat{f}_L=0$.
In the hydrodynamic limit ($q \rightarrow 0$) one obtains the complex conjugated eigenfrequencies
\begin{equation}
    \omega_{\pm}(\textbf{q}) = c_{\pm}(\theta_q)q-i\epsilon_{\pm}(h, \textbf{q})\,.
    \label{norm_freq}
\end{equation}
Their real parts (the sound speeds) are unaffected by the external field and are given by
\begin{equation} \label{cpm}
    c_{\pm}(\theta_q)= \bigg( \frac{\gamma+v_2}{2}\bigg) \cos(\theta_q) \pm c_2(\theta_q);
\end{equation}
with
\begin{equation}\label{c2}
    c_2(\theta_q)\equiv \sqrt{\frac{(\gamma-v_2)^2\cos(\theta_q)^2}{4}+c_0^2\sin(\theta_q)^2}\,.
\end{equation}
The only field-dependent terms are found to be the imaginary dampings $\epsilon_\pm(h, {\bf q})$ equal to
\begin{equation}
\epsilon_\pm (h, {\bf q}) = \epsilon_\pm (0, {\bf q})+ a_\pm(\theta_q) h \equiv b_\pm(\theta_q) q^2 + a_\pm(\theta_q) h\,,
\end{equation}
where $a_\pm(\theta_q)$ is given by Eq. \eqref{eq:apm} and 
\begin{equation}\label{A3}
b_\pm(\theta_q)=\frac{\mathrm{\Xi}_\pm(\theta_q)}{[2c_{\pm}(\theta_q)-(v_2+\gamma)\cos(\theta_q)]}\,.
\end{equation}
The numerator $\Xi_\pm$ of Eq. \eqref{A3} is rather complicated but only depends on the angle $\theta_q$,
\begin{equation}
\begin{split}
&\Xi_\pm(\theta_q) = 
-[D_L\sin^2(\theta_q) + D_\parallel\cos^2(\theta_q)]v_2\cos(\theta_q)+\\
&+[D_L \sin^2(\theta_q) + D_\parallel \cos^2(\theta_q) -\phi c_\pm(\theta_q)\,\cos(\theta_q)] 
c_\pm(\theta_q)+\\
&-[D_{\rho\parallel}\!\cos^2(\theta_q)\!+\!D_{\rho\perp}\!\sin^2(\theta_q)\!-\!\phi c_\pm(\theta_q)\!\cos(\theta_q)]\gamma\!\cos(\theta_q)
\\&+\frac{c_0^2}{\rho_0}D_{\rho v}\!\cos(\theta_q)\!\sin^2(\theta_q)\!-\![g_t c_\pm(\theta_q)\!+\! g_\parallel\!\cos(\theta_q)]\rho_0\!\sin^2(\theta_q).
\end{split}
\end{equation}

The solution of Eqs.\eqref{rho_four_first}-\eqref{v_L_first} can be now easily expressed in terms of the above eigenfrequencies $\omega_\pm$ (the zeros of the associated matrix determinant). \\
In particular, in the hydrodynamic limit ($q\to0)$ we have
\begin{equation}
\delta\hat{\rho}(\omega, {\bf q})=\frac{\rho_{\rm 0}q\sin(\theta_q) \hat{f}_L}{(\omega-\omega_+({\bf q}))(\omega-\omega_-({\bf q}))}\,.  
\end{equation}
Correlating this solution pairwise we obtain 
\begin{widetext}
\begin{equation} \label{auto_corr_dens}
        \langle \delta\hat{\rho}(\textbf{q},\omega) \delta\hat{\rho}(-\textbf{q},-\omega) \rangle
        = \frac{\rho_0^2 q^2 \sin^2(\theta_q) \Delta}{\{ [\omega-c_+(\theta_q) q]^2+[\epsilon_+(h, {\bf q})]^2\} \{[\omega-c_-(\theta_q) q]^2+ [\epsilon_-(h, {\bf q})]^2\}}.
\end{equation}
\end{widetext}
The equal time structure factor \eqref{autocorr_eqt} is finally recovered transforming back in real (equal) time by an integration over $\omega$.

\bibliography{bibliografia}
\end{document}